
\magnification=1200
\hsize=14.4truecm
\vsize=23.1truecm
\baselineskip=17pt plus 1pt minus 1pt
\parindent=1truecm
\centerline {\bf INFORMATION LENGTH AND LOCALIZATION IN ONE DIMENSION}
\bigskip
\centerline{Imre Varga and J\'anos Pipek}
\bigskip
\centerline {\it Quantum Theory Group, Institute of Physics,}
\centerline {\it Technical University of Budapest,
H--1521 Budapest, Hungary}
\bigskip
    {\bf Abstract} - The scaling properties of the wave functions in
    finite samples of the one dimensional Anderson model are analyzed.
    The states have been characterized using a new form of the
    information or entropic length, and compared with analytical
    results obtained by assuming an exponential envelope function. A
    perfect agreement is obtained already for systems of
    $10^3$--$10^4$ sites over a very wide range of disorder parameter
    $10^{-4}<W<10^4$. Implications for higher dimensions are also
    presented.
\bigskip

\bigskip
\bigskip

    The numerical detection of exponential localization in finite,
    random systems is not a trivial task, especially in the weak
    disorder limit when the localization length of the eigenstates is
    expected to be larger than the system size. This may
    be a problem even in one dimension (1D), where rigorous results
    [1] affirm complete exponential localization for any strength of
    disorder.

    The model under consideration can be described by a tight--binding
    Schr\"o\-dinger equation in the nearest neighbor approximation as
    $$u_{n+1}+u_{n-1}+V_nu_n=Eu_n,                          \eqno(1)$$
    where $V_n$ are independent random variables with uniform
    distribution over the $[-W/2\dots W/2]$ interval, $E$ is the
    eigenenergy and $u_n$ is the amplitude of the eigenfunction on
    site $n$. The wave functions are expected to behave asymptotically
    up to oscillations as
    $$u_n\sim \exp(-\gamma n),                              \eqno(2)$$
    where $\gamma=\xi^{-1}$ is the inverse localization length or
    Lyapunov exponent that may be numerically obtained as [1]
    $$\gamma\approx\gamma_N(E)=
      {1\over N}\sum_{i=1}^N\ln\left |{u_n}\over{u_{n-1}}\right |.
                                                            \eqno(3)$$

    The exponential localization has been corroborated by the
    one--parameter scaling theory [2]. This was built essentially on
    the basis of the Thouless number [3] that is related to the
    dimensionless conductance [4]. Inspired by that scaling law in a
    recent paper Casati {\it et al.} [5] have introduced the concept
    of {\it information localization length} [6] and investigated
    numerically the possible scaling properties of the eigenstates
    themselves. Their study has also been motivated by previous
    results obtaining a scaling law for an analogous model in quantum
    chaos, the kicked rotator [7], as well as for band random matrices
    [8] using the same concept.

    The main idea of Ref. [5] is to take a suitable ensemble of
    eigenstates and to calculate the information length as
    $$\beta_C (E,N,W)=\exp (\overline S - S_{ref}),         \eqno(4)$$
    where $N$ stands for the system size, $W$ describes the strength
    of disorder. Index $C$ is used to label $\beta$ in order to refer
    to the definition of Casati {\it et al.} $\overline S$ is the
    averaged information entropy,
    $$S = -\sum_{n=1}^N u_n^2\ln u_n^2,                     \eqno(5)$$
    of normalized eigenstates in a window around energy $E$ for
    different realizations of the random potential.
    The $S_{ref}$ in Eq. (1) stands for the entropy of a reference
    state
    $$u_n\sim\sin(\varphi n)\qquad {\rm where}\qquad \cos\varphi =E/2.
                                                            \eqno(6)$$
    This wave function is the exact solution of Eq. (1) in the absence
    of disorder ($W=0$) with $u_0=0$ and $u_1=1$. A straightforward
    calculation yields the asymptotic form $S_{ref}(N)\to\ln(2N)-1$
    as $N\to\infty$. We have to indicate that the particular choice of
    $S_{ref}$ in Eq. (4) involves a delicate problem, that we wish to
    discuss below.

    The principal aim of definition (4) is that $\beta_C$ should give
    the portion of the sites significantly populated by the
    eigenstates compared to that of the reference state (6). It is
    clear that with the increase of the system size $N$ we expect
    $\exp (S)\propto N$ in the case of extended states and $\exp
    (S)\to $~const. for localized states. Note, that since for large
    $N$, $\exp (-S_{ref})\sim c/N$, its role is normalization.

    Casati {\it et al.} have numerically established the scaling law
    [5]
    $$\ln{{\beta_C}\over {1-\beta_C}}=
                  \ln\left ({\xi_{\infty}\over N}\right) + C\eqno(7)$$
    with $C\approx 1$. $\xi_{\infty}$ has been calculated as
    $\xi_{\infty}=1/\gamma_N$ (cf. Eq. (3)) for strong disorder
    ($1/\gamma_N\leq N$). For weak disorder ($1/\gamma_N>N$) instead
    of numerically calculating the Lyapunov exponent the authors of
    Ref. [5] used $\xi_\infty$ given by the perturbative calculation
    [9]. They have found, however, no theoretical explanation for (7).

    We would like to point out that the approach of Casati {\it et
    al.} [5] resides on the supposition that in Eq. (7) $\beta_C<1$
    i.e. $\overline S < S_{ref}$. There is, however, no rigorous proof
    for these relations. This is particularly crucial in the weak
    localization limit $W\to 0$, $\beta_C\approx 1$. Indeed, in actual
    numerical calculations for small ensembles we
    have found occasionally $\beta_C>1$, as well. We could prove,
    however, that $\overline S\leq S_{ref}(N)$ for sufficiently large
    ensembles. The proof is based on the idea that any small
    perturbation $\{Q_n^\prime=Q_n(1+\delta_n)\}$,
    $\overline\delta_n=0$ of an arbitrary probability distribution
    $\{Q_n\}$ leads to $\overline S^\prime\leq S$, where $\overline
    S^\prime$ is an average over an infinite number of realizations of
    $\delta_n$. As a byproduct, it is also evident that for the
    calculation of $\beta_C$, $S_{ref}=S_{ref}(N)$ has to be applied
    instead of using the asymptotic form $\ln (2N)-1$.

    Furthermore, as we will show it later, for finite systems the
    functional form of the scaling law (7) does not seem to be valid
    in the weak localization limit ($W\to 0$). This is mainly due to
    the improper application of the perturbative treatment.

    In this Letter we propose a new form of $\beta$ for resolving
    the above mentioned difficulties and appropriately handling the
    $W\to 0$ limit. We will demonstrate that our $\beta$ function is,
    apart from showing special scaling properties, capable to prove
    the exponential localization in one dimension.

    On the basis of our recently introduced classification scheme
    [10], the information entropy (5) of any general, normalized,
    nonnegative lattice distribution can be split as a sum of two
    terms
    $$S=S_{str}+\ln D,                                      \eqno(8)$$
    where $D$ is the delocalization measure [11] or participation
    number [12]
    $$D=\left (\sum_{n=1}^N u_n^4\right )^{-1},             \eqno(9)$$
    and $S_{str}$ is the {\it structural entropy} of the distribution.
    Parameter $D$ is widely used in the literature giving the number
    of sites the eigenstate extends to. Therefore it is bounded as
    $1\leq D\leq N$. Using $D$ we may introduce a normalized quantity
    $q$, the spatial filling factor or participation ratio as
    $$q=D/N,\qquad {\rm for\ which}\qquad 0<q\leq 1.       \eqno(10)$$
    The structural entropy in equation (8) has been shown [10] to be
    nonnegative with bounds
    $$0\leq S_{str}\leq -\ln q.                            \eqno(11)$$

    Using the quantities discussed above we now propose an alternative
    form of the normalized information length of an eigenstate as
    $$\beta={1 \over N}\exp S,                             \eqno(12)$$
    which using expressions (8) (9) and (10) becomes
    $$\beta=q\exp (S_{str}).                               \eqno(13)$$
    Due to the well-known properties $0\leq S\leq\ln N$ the following
    restrictions are imposed on $\beta$
    $$0<\beta\leq 1.                                       \eqno(14)$$
    These bounds are valid for each state separately, whereas
    $\beta_C\leq 1$ can only be guaranteed after an appropriate
    averaging process.
    Obviously, an eigenstate expanding uniformly over the whole system
    will have $u_n^2=1/N$ so that $\beta=1$. In the other extreme, for
    localized states $D\sim 1$ and $S_{str}\approx 0$, in a finite
    system one obtains $\beta\approx 1/N$.

    Returning now to the Anderson model (1) with non-zero disorder
    $W\neq 0$, one expects according to (2) and (6) the charge
    distribution of the solution of the form
    $$|u_n|^2\sim f(\gamma n)\sin^2(\alpha n),             \eqno(15)$$
    where $f$ is a slowly varying envelope function due to the
    presence of the perturbing random potential. Obviously, in our
    case function $f$ is expected to take an exponential form $f(\rho
    )=\exp (-\rho)$. The value of $\alpha$ is roughly $\varphi$
    defined
    in (6). Further specification of $\alpha$ is needless for our
    purposes, the only restriction we impose is $\alpha\gg 1/N$ that
    is always fulfilled except very close to the bandedges. As we have
    shown [10,13] for multiplicative superstructures of the form (15)
    we get
    $$\ln q=\ln q^f +\ln q^0                               \eqno(16)$$
    and
    $$S_{str}=S_{str}^f + S_{str}^0,                       \eqno(17)$$
    where the upper index $f$ stands for the values obtained for the
    charge distribution $f(\gamma n)$ alone and upper index $0$
    indicates the ones for $\sin^2(\alpha n)$. It is possible to show
    that (independently of $\alpha$)
    $$q^0=2/3 \qquad {\rm and} \qquad S_{str}^0=\ln 3-1.   \eqno(18)$$
    Using (16) and (17) as well as definition (13) we get
    $$\beta =\beta_f\,\beta_0,                             \eqno(19)$$
    where $\beta_0=q^0\exp(S_{str}^0)\approx 0.7357$. In the limit of
    vanishing disorder $W\to 0$ one expects $\gamma\to 0$, $f(\rho
    )\to 1$. For such distribution using (5), (8) and (9) $q^f\to 1$
    and $S_{str}^f\to 0$ therefore $\beta_f\to 1$. For strong disorder
    on the other hand $\gamma\gg 1$ which yields $D\approx 1$,
    therefore in finite systems $q^f\approx 1/N$ and $S_{str}^f\approx
    0$ resulting $\beta_f\approx 1/N$.

    The role of $\beta_0$ in (19) is similar to the factor $\exp
    (S_{ref})$ in (4) introduced by Casati {\it et al.} [5], however,
    $\beta_0$ is a constant independent of the system size $N$, and
    its derivation is based on the separation of the wave function to
    an envelope and a strongly oscillating part (15).

    Before performing the numerical simulation we still have to give
    the explicit $q^f$ and $S_{str}^f$ values as a function of
    $\gamma$. In Ref. [10] the general form of $q^f(z)$ and
    $S_{str}^f(z)$ functions with $z=\gamma N=N/\xi$ is given for
    arbitrary dimensionality, that have been calculated applying a
    continuous lattice approximation, i.e. the relevant scale
    ($\gamma^{-1}$) was assumed to extend over many lattice spacings.
    It has been shown in [10] that in one dimension
    $$q^f(z)={{[F(z)]^2} \over {z\,G(z)}},          \eqno(20{\rm a})$$
    and
    $$S_{str}^f(z)={{H(z)}\over{F(z)}}+
                 \ln\left({{G(z)}\over{F(z)}}\right),
                                                    \eqno(20{\rm b})$$
    where functions $F(z)$, $G(z)$, and $H(z)$ are defined as
    $$F(z)=\int_0^z\,f(\rho )\,d\rho ,              \eqno(21{\rm a})$$
    $$G(z)=\int_0^z\,f^2(\rho )\,d\rho ,            \eqno(21{\rm b})$$
    $$H(z)=-\int_0^z\,f(\rho )\ln [f(\rho )]\,d\rho .
                                                    \eqno(21{\rm c})$$
    Inserting the general functions given in expressions (21) into
    (20) one obtains for $\beta_f$ in the continuous limit
    $$\beta_f(z)=q^f(z)\exp (S_{str}^f(z))=z^{-1}\,F(z)\,
                 \exp\left({{H(z)}\over{F(z)}}\right).     \eqno(22)$$
    It is straightforward to calculate the $\beta_f(z)$ function for
    any envelope shape $f(\rho )$. For exponential decay, $f(\rho
    )=\exp (-\rho )$, we get
    $$\beta_{\rm exp}(z)={{\exp z-1}\over {z\,\exp z}}
                 \exp\left (1-{z\over {\exp z-1}}\right ). \eqno(23)$$
    Expressions (22) and (23) are the principal results of this
    Letter showing the scaling property of $\beta_f$ provided that a
    reasonable definition for the $f(\rho )$ decay function (15)
    exists.

    Let us turn now to the asymptotic properties of $\beta_f(z)$.
    In the case of strong localization $z\to\infty$ ($N\gg\xi$),
    since both $F(\infty )$ and $H(\infty )$ are finite for most of
    the practical cases, from the general expression (22) one gets
    $$\beta_f(z)\sim z^{-1}\sim {\xi\over N},              \eqno(24)$$
    as expected. In the other limit of delocalization as $z\to 0$
    (e.g. $\xi\to\infty$ keeping $N$ fixed) we found that the
    asymptotic form of $\beta_f(z)$ is governed by the short range
    properties of the form function $f(\rho )$, i.e. it depends on the
    derivative of $f(\rho )$ at the origin $\rho =0$. Namely, if
    $f^{\prime}(0)\neq 0$ then
    $$\beta(z)\approx 1-{1\over 24}z^2,                    \eqno(25)$$
    while for a Gaussian form function e.g. where $f^{\prime}(0)=0$
    the first nonvanishing term is of the order of $z^4$.

    Instead of the $\beta_f(z)$, after Ref. [5], we define
    $$y(z)={{\beta_f(z)}\over {1-\beta_f(z)}},             \eqno(26)$$
    in order to emphasize both the localized and delocalized limits.
    In our numerical simulation we have compared Eq. (26) for
    exponential form function (22) with the calculated
    $$y={{\beta_f}\over {1-\beta_f}}                       \eqno(27)$$
    values, where $\beta_f$ is defined here as
    $$\beta_f={1\over {\beta_0}}\overline q\,\exp\overline S_{str}.
                                                           \eqno(28)$$
    As $0<\beta_f\leq 1$ is true already for individual wave functions
    it is not necessary to perform averaging over states taken from an
    energy window, however, at a certain energy we calculate the
    statistical means $\overline q$ and $\overline S_{str}$ over many
    realizations of the random potential.

    The eigenvectors in our simulation were obtained by the iteration
    of the recurrence relation of Eq. (1) with initial conditions
    $u_0=0$ and $u_1=1$. In all of the presented results the length of
    the system was $N=10^4$ and the number of samples used for
    averaging was $M=10^3$. The energy was fixed to $E=0.1$. The
    localization length was obtained as
    $$\xi(E) = \overline\gamma(E)^{-1},                    \eqno(29)$$
    were $\gamma(E)$ was calculated according to Eq. (3). In a finite
    lattice two relevant length scales characterize the system the
    chain length $N$ and the lattice spacing ($a=1$). The relation of
    $\xi$ with respect to these length scales is essential in such
    type of calculations.

    In Figure 1 we have plotted the localization length as a
    function of the strength of disorder ranging from $W=10^{-4}$ up
    to $W=10^4$. Our numerical calculations confirm the theoretically
    expected behavior $\xi^{-1}\sim\ln W$ for large disorder. In the
    case of vanishing disorder $W\to 0$ perturbation theory predicts
    $\xi^{-1}\sim W^2$ in the thermodynamic limit $N\to\infty$ [9].
    This consideration, however, fails for finite $N$ as one can see
    in the low $W$ part of Figure 1, where we have numerically found
    $\xi^{-1}\sim W$. The figure clearly shows that this finite size
    behavior becomes relevant for such disorder values where the
    localization length is comparable to or greater than the system
    size. It seems that the behavior of the wave function on
    intermediate length scales is governed by a different
    characteristic length $\xi_i$. As is shown in Figure 2 on this
    latter scale exponential localization can be detected, as well.

    In Figure 2 we have compared $\beta_C$ of Casati {\it et al.} and
    our numerical $\beta_f$ and theoretical $\beta_{\rm exp}$ as a
    function of $\xi/N$. In each of these cases we have plotted
    $y=\beta/(1-\beta)$ vs $\xi/N$ in a log-log plot.
    The results of the simulation follow the
    curve for exponential localization. The deviation for
    $\ln(\xi/N)\leq -9$ is present because the value of the
    localization length becomes comparable to the lattice constant
    as $W\to\infty$ i.e. we obtain $\ln y\to -\ln N$.
    Both our analytical and numerical results confirm the expected
    high disorder behavior of $\ln y\approx\ln\beta\sim \ln\xi$ (see
    e.g. (24)), as well as the low disorder behavior of $\ln y\approx
    \ln(1-\beta)\sim 2\ln\xi$ (see e.g. (25)). For strong disorder we
    observe perfect agreement with the scaling law set up by Casati
    {\it et al.} [5] in Eq. (7). Our analytical results (23) confirm
    this scaling law
    $$\ln\beta_{\rm exp}(z)\to -\ln z+1,                  \eqno(30)$$
    in the limit of strong localization $z=N/\xi\to\infty$.

    For weak disorder, however, Figure 2 shows a considerable
    disagreement between our results and that of Ref. [5]. This is
    easy to understand considering that for this regime Casati {\it et
    al.} have used the predictions of the perturbative calculation
    valid in the thermodynamic limit. As we pointed out earlier this
    approach needs a careful analysis. They compare quantities $y$ and
    $\xi_{\infty}$ where $y$ is calculated from wave functions
    characterized by the intermediate length scale $\xi_i$. On the
    other hand calculating the $\xi$ according to Eqs. (3) and (29) we
    obtain an almost perfect agreement between the numerical
    simulation and our analytical expressions for exponential form
    function. This shows that, apart from the exponential long range
    behavior for $\xi_{\infty} < N$, in the intermediate range
    ($\xi_{\infty} > N$) the same kind of decay was found with a
    different scale constant $\xi_i$, as well.

    Just to have a feeling how well the charge distribution of the
    form of Eq. (15) describes the average properties of the wave
    functions in the Anderson model, in Figure 3 we have plotted
    $\overline S_{str}$ as a function of $\overline q$. We have
    compared the results of the simulation with analytical results
    obtained assuming the form of Eq. (15). A satisfactory agreement
    can be established between the numerical and analytical results
    especially for low and high disorder. The charge distribution is
    clearly not of pure exponential form, but a plane wave modulated
    by an exponential envelope.

    We would like to note that especially for weak disorder, the
    exponential localization, apart form fluctuations and
    oscillations, is still strictly true and visible using our
    construction of $\beta$--function at least up to localization
    lengths several times larger than the size of the system. We
    believe that after a proper definition of the localization length,
    a similar procedure could clearly show the expected exponential
    localization in two dimensions, as well. Results along these lines
    are to be published in a subsequent paper.

\bigskip
    Financial support from Orsz\'agos Tudom\'anyos Kutat\'asi Alap
    (OTKA), Grant No. 517/1991 and T7283/1993 is gratefully
    acknowledged.
\vfill\eject
\baselineskip=.8truecm plus .1truecm
    {\bf References}
\medskip
\leftskip 1truecm

\item{[1]} Ihsii K 1973 {\it Prog. Theor. Phys. Suppl.}, {\bf 53}
    77; Kunz H and Souillard B 1980 {\it Commun. Math. Phys.}, {\bf
    78} 201; Delyon F, Levy Y and Souillard B 1985 {\it Phys. Rev.
    Lett.}, {\bf 55} 618

\item{[2]} MacKinnon A and Kramer B 1981 {\it Phys. Rev. Lett.} {\bf
    47} 1546

\item{[3]} Licciardello D C and Thouless D J 1975 {\it J. Phys. C:
    Solid State Phys.} {\bf 8} 4157; $ibid$ 1978 {\bf 11} 925

\item{[4]} Abrahams E, Anderson P W, Licciardello D C and
    Ramakrishnan T V, 1979 {\it Phys. Rev. Lett.} {\bf 42} 673

\item{[5]} Casati G, Guarneri I, Izrailev F, Fishman S, and Molinari
    L 1992 {\it J. Phys.: Condens. Matter} {\bf 4} 149

\item{[6]} Izrailev F M 1990 {\it Phys. Rep.} {\it 196} 299

\item{[7]} Casati G, Guarneri I, Izrailev F, and Scharf R 1990 {\it
    Phys. Rev. Lett.} {\bf 64} 5

\item{[8]} Casati G, Molinari L, and Izrailev F 1990 {\it Phys. Rev.
    Lett.} {\bf 64} 1851

\item{[9]} Derrida B and Gardner E 1984 {\it J. Physique} {\bf 45}
    1283.

\item{[10]} Pipek J and Varga I 1992 {\it Phys. Rev.} {\bf A46} 3148

\item{[11]} Pipek J 1989 {\it Intern. J. Quantum Chem.} {\bf 36}
    487; and Varga I, to be published

\item{[12]} Thouless D J 1974 {\it Phys. Rep.} {\bf 13} 93

\item{[13]} Pipek J and Varga I 1993 {\it Intern. J. Quantum Chem.},
    in press.

\vfill \eject
    {\bf Figure Captions}
\medskip
\parindent -1.5truecm
\leftskip 1.5truecm
\hsize=12.5truecm
    {\bf Figure 1.}\ The log-log plot of the localization length
    versus strength of disorder using the numerical simulation. The
    characteristic relations are also denoted.
\bigskip
    {\bf Figure 2.}\ Scaling of the entropic length versus the
    localization length using $\ln y=\ln (\beta/(1-\beta))$
    versus $\ln (\xi/N)$. Solid symbols represent the results of our
    numerical simulation. Dashed line stands for the scaling law found
    numerically by Casati {\it et al.} [5]. Note the change in
    the slope of the continuous curve at around $\xi\approx N$.
    The continuous curve is our analytical result (see Eqs.
    (23) and (26)).
\bigskip
    {\bf Figure 3.}\ The structural entropy versus filling factor in a
    semi-log plot. The filled symbols represent our simulation, while
    the solid line stands for the relation assuming a charge
    distribution of the form of Eq. (15).
\bye